\documentclass[twocolumn,amsmath,amssymb,
aip,
10pt,showkeys,superscriptaddress,longbibliography
,nofootinbib
]{revtex4-2}
\usepackage{dcolumn}
\usepackage{bm}
\usepackage{mathtools, cases}
\usepackage[unicode]{hyperref}
\usepackage{graphicx}
\usepackage{times}
\usepackage{xcolor}
\hypersetup{colorlinks=true,linkcolor=blue,citecolor=blue,filecolor=black,urlcolor=blue}

\begin{document}

\title{Dense random packing with a power-law size distribution: the structure factor, mass-radius relation, and pair distribution function
}

\author{Alexander Yu. Cherny}
\email[Corresponding author, e-mail:~]{cherny@theor.jinr.ru}
\affiliation{Joint Institute for Nuclear Research, Dubna 141980, Russian Federation}

\author{Eugen M. Anitas}
\affiliation{Joint Institute for Nuclear Research, Dubna 141980, Russian Federation}
\affiliation{Horia Hulubei, National Institute of Physics and Nuclear Engineering, RO-077125 Bucharest-Magurele, Romania}

\author{Vladimir A. Osipov}
\affiliation{Joint Institute for Nuclear Research, Dubna 141980, Russian Federation}

\date{\today}

\begin{abstract}
We consider dense random packing of disks with a power-law distribution of radii and investigate their correlation properties. We study the corresponding structure factor, mass-radius relation and pair distribution function of the disk centers. A toy model of dense segments in one dimension (1d) is solved exactly. It is shown theoretically in 1d and numerically in 1d and 2d that such packing exhibits fractal properties. It is found that the exponent of the power-law distribution and the fractal dimension coincide. An approximate relation for the structure factor in arbitrary dimension is derived, which can be used as a fitting formula in small-angle scattering. The findings can be useful for understanding microstructural properties of various systems like ultra-high performance concrete, high-internal-phase ratio emulsions or biological systems.
\end{abstract}


\maketitle

\section{Introduction}
\label{sec:intro}

Models of dense packing play an important role in soft matter, chemistry and material science. They are important for describing the physical and chemical properties of colloids, biological systems, granular materials, glasses, and many other systems (see the review \cite{rev:Torquato18}). In particular, minimisation of porosity in a system of polydisperse particles directly relates to the problem of creating ultra-high performance concrete \cite{concrete04}. \emph{Dense packing} of polydisperse particles are important in high internal-phase-ratio emulsions~\cite{kwok20}, where the coalescence mechanisms in concentrated emulsions changes as compared to diluted ones, in sandstones, where nonwetting phase clusters (ganglia) follow a power-law size distribution~\cite{iglauer10}, etc. Also, at high packing fractions, equilibrium mixtures consisting of large and small spheres can separate, and the disorder-order phase transition observed in monodisperse packings can be suppressed by a small degree of polydispersity~\cite{torquato10}. In living systems, a power-law distribution of molecule abundance levels is known to be a robust characteristic, regardless of the time and space~\cite{sato18}.

In spite of the simple formulation, dense packing remains a long-standing problems in mathematics, not yet completely solved \cite{Conway99:book,torquato09}. Packing of polydisperse particles \cite{santiso02,raoush07,farr09} is one of the most difficult: for a given distribution of sizes, it is difficult to say in general whether the complete packing is attainable and how to optimize it \cite{Voivret07,Reis12}. The densest packing of polydisperse particles is known to be reachable with power-law size distributions~\cite{Reis12}. However, very little is known about the characteristics of such systems. In particular, the spatial correlations between particle centers or the density fluctuations are unknown.

In this paper, we study the spatial correlations between positions of the centers of densely packed $d$-dimensional spheres, whose radii are randomly distributed in accordance with a power-law. We restrict ourselves to one and two dimensions. The number of spheres $d N(r)$ whose radii fall within the range ($r,r + d r$) is proportional to $dr/r^{D+1}$. The exponent obeys the condition $0<D<d$ and the radii vary from $a$ to $R$. Here $d$ is the Euclidian dimension of space. Aste has shown \cite{Aste96} that the full packing is possible when the exponent $D$ lies between the Hausdorff dimension of the Apollonian packing and $d$. The dimension of the Apollonian packing amounts to $D_\mathrm{Ap}=1.3057\ldots$ \cite{manna91} and $D_\mathrm{Ap}=2.4739\ldots$ \cite{borkovec94} in two and there dimensions, respectively. However, the statement of Aste has no solid mathematical justification and is considered as a conjecture \cite{Botet_21}. On the other hand, the full packing is definitely reachable if $D$ is sufficiently close to $d$ \cite{Botet_21}. In one dimension, $D$ has no restrictions from below and the full packing of a set of segments can always be realized, then $D$ can vary from zero to one.

The usual approach to a system of spheres is to apply the Gibbs statistical mechanics with different approximations (like the Ornstein–Zernike equation with  the Percus–Yevick approximation) \cite{Salacuse82,Lado96,Botet21a}. However, the Gibbs statistical mechanics is questionable for the full dense packing, since the phase space $(\textbf{p},\textbf{r})$ degenerates to one point: $\textbf{p}=0$ and $\textbf{r}$ being equal to static positions of $N$ particles.

In 1d, we develop a direct probability approach with the infinite limit of filling particles. The peculiarity of the limit is related to the nature of dense packing: the system volume remains constant while the number of particles tends to infinity. By contrast, in the usual thermodynamic limit, both the number of particles and volume tend to infinity.

We study numerically the densely packed segments in 1d and disks in 2d. It is shown that such systems exhibits fractal-like properties  (see, e.g., Refs.~\cite{Cherny2017SmallangleSnowflake,cherny19rev}). We found that the mass-radius relation for the $d$-sphere centers of unit weight is proportional to $r^D$ as for real fractals, provided the system is densely packed. This implies that the exponent of the power-law distribution and the fractal dimension coincide. The mass-radius relation $M(r)$ is intimately related to the pair distribution function $g(r)$ and the structure factor $S(q)$ \cite{cherny11}. The latter can be measured directly, say, in small-angle scattering experiments at nano- and micro-scales. Once the full packing is reached then $M(r)\sim r^D$, $g(r)\sim r^{D-d}$, and $S(q)\sim q^{-D}$. In 1d, the theory matches very well the numerical simulations, in 2d we have only numerical results yet.

We start in Sec.~\ref{sec:MrSqGen} with the relevant theoretical background needed for the rest of the paper. Section \ref{sec:Distr_pack} discusses the important question of the limit of infinite number of particles in dense packing. A one-dimensional exactly solvable model is considered in Sec.~\ref{sec:entireDim}, where the structure factor, mass-radius relation, and the pair distribution function for a set of dense segments are found both analytically and numerically. The same correlation properties are then obtained numerically for 2d dense packed disks in Sec.~\ref{sec:2d}. Further, Sec.~\ref{sec:empir} provides an approximate relation for the structure factor in arbitrary dimension. Finally, in Sec.~\ref{sec:concl} we summarize the obtained results and present some prospects for future research.

\section{The structure factor, pair distribution function, and mass-radius relation for a set of points}
\label{sec:MrSqGen}

For a set of $N$ points of unit weight located at the positions $\textbf{r}_{1},\cdots,\textbf{r}_{N}$, the mass-radius relation is defined as the average value of mass $M(r)$ enclosed in the imaginary sphere of radius $r$, which is centered on a point belonging to the set \cite{Gouyet1996PhysicsStructures}. According to the definition, it is given by
\begin{align}
 M(r)=\frac{1}{N}\sum_{i,j}\theta(r-r_{ij})=1+\frac{1}{N}\sum_{i\neq j}\theta(r-r_{ij}), \label{Mrdef}
\end{align}
where $r_{ij}=|\textbf{r}_{i}-\textbf{r}_{j}|$ and  $\theta(z)$ is the Heaviside step function, that is, $\theta(x)=1$ for $x\geqslant0$ and zero elsewhere. Then $M(r)=1$ when $r$ is less than the smallest distance between points and $M(r)=N$ when $r$ exceeds the largest distance.

We define the structure factor of the set of points as \cite{cherny11}
\begin{align}\label{Sqdef}
S(q)=\frac{1}{N}\left\langle\rho_{\textbf{q}}\rho_{-\textbf{q}}\right\rangle_{\hat{q}},
\end{align}
where $\rho_{\textbf{q}}=\sum_{j}e^{-i\textbf{q}\cdot\textbf{r}_{j}}$ is the Fourier transform of the density of the points $\rho(\textbf{r}) =\sum_{j} \delta(\textbf{r}-\textbf{r}_{j})$, and the brackets $\langle\cdots\rangle_{\hat{q}}$ stand for the average over all directions of unit vector $\hat{q}$ along $\textbf{q}$. By definition, the structure factor depends only on the absolute value of $\textbf{q}$. It has the asymptotics $S(q)\simeq 1$ when $q\to\infty$ and $S(q)=N$ at $q=0$. The structure factor can be measured in small-angle scattering experiments.

The pair distribution function is given by
\begin{align}
g(r)=&\frac{V}{N(N-1)}\sum_{i\neq j}\left\langle\delta(\textbf{r}-\textbf{r}_{i} +\textbf{r}_{j})\right\rangle_{\hat{r}}\nonumber\\
=&\frac{V}{N(N-1)}\frac{1}{\Omega_{d} r^{d-1}}\sum_{i\neq j}\delta(r-r_{ij}).\label{grdef}
\end{align}
Here $V$ is the $d$-dimensional volume of the system, $\Omega_{d}=2\pi^{d/2}/\Gamma(d/2)$ is the area of unit sphere in $d$ dimension, and $\Gamma(s)$ is the gamma function. The pair distribution function describes the spatial correlations between points and is proportional to the probability density to find a particle at the distance $r$ from another particle.

One can also introduce the probability density of finding the distance $r$ between two arbitrarily taken points
\begin{align}\label{pdef}
p(r)=\frac{1}{N(N-1)}\sum_{i\neq j}\delta(r-r_{ij})=\frac{\Omega_{d} r^{d-1}}{V} g(r).
\end{align}
This quantity is called the pair distance distribution function. It is widely used in the theory of small-angle scattering (see, e.g., Ref.~\cite{pedersen99}). As it follows from the definition (\ref{pdef}), the pair distance distribution function obeys the normalization condition $\int_{0}^{\infty}dr\,p(r) =1$. The last equality in (\ref{pdef}) relates $p(r)$ to the pair distribution function.

All the above quantities are intimately connected to each other:
\begin{align}
g(r)\frac{N-1}{V}=&\frac{1}{(2\pi)^{d}}\int_{0}^{\infty} dq\,\Omega_{d} q^{d-1}f_{d}(q r) [S(q)-1]\nonumber\\
=&\frac{1}{\Omega_{d} r^{d-1}}\frac{\partial M}{\partial r}, \label{relg_S_M}
\end{align}
where the function $f_{d}(z)$ is given by
\begin{align}
f_{d}(z)=\frac{\Gamma({d}/{2})J_{d/2-1}(z)}{(z/2)^{d/2-1}}=\begin{cases}
\cos z, &d=1,\\
J_{0}(z),&d=2,\\
\sin z/z,&d=3,
\end{cases}\label{fzdef}
\end{align}
where $J_{n}(z)$ is the Bessel function of $n$th order. Note that the pair distribution function becomes zero when the distance exceeds the maximum separation between centers, since $M(r)$ is a constant there. By contrast, for a usual thermodynamic system, $M(r)\simeq \frac{N}{V}\frac{\Omega_{d}}{d}r^d$ at sufficiently big distances, which yields $g(r)\simeq 1$ from Eq.~(\ref{relg_S_M}). For the same reason, $g(0)=0$ in the case of finite $N$.

One can directly relate the structure factor and mass-radius relation by integrating Eq.~(\ref{relg_S_M}) with respect to $r$:
\begin{align}\label{relM_S}
M(r)=1+\frac{\Omega_{d}^{2}}{(2\pi)^{d}}\int_{0}^{\infty} dz\,\frac{h_{d}(z)}{z}\left[S(z/r)-1\right].
\end{align}
Here we denote
\begin{align}
h_{d}(z)= \frac{\Gamma({d}/{2})z^{d/2}J_{{d}/{2}}(z)}{2^{1-{d}/{2}}}=\begin{cases}
\sin z, &d=1,\\
z J_{1}(z),&d=2,\\
\sin z-z\cos z,&d=3.
\end{cases}
\label{gzdef}
\end{align}
The relations (\ref{relg_S_M})-(\ref{gzdef}) were obtained in the three-dimensional case in Ref.~\cite{cherny11}.

\section{The distribution of radii in the limit of dense packing}
\label{sec:Distr_pack}

Let us consider a set of $N$ non-overlapping disks, which completely fill a square of fixed size in the limit $N\to\infty$. We denote the number of disks, whose radii are bigger or equal to $r$, by $N(r)$. If $R$ and $a$ are the largest and smallest radii, respectively, then $N(R)=1$ and $N(r)=N$ when $r\leqslant a$. It follows that the probability distribution of radii, proportional to $-dN(r)/dr$, is cut from below at $r=a$ for a finite number of disks $N$.  In the case of a power-law distribution with the exponent $0<D<d$, we have $N(r)\sim 1/r^D$ for $a \leqslant r\ll R$.

We observe that once the dense packing is realized for an arbitrary probability density, its cumulative distribution $N(r)$ should be normalized so that $N(R)=1$ in the limit $N\to\infty$. For the power-law distribution, it follows that $N=N(a)\sim 1/a^D$, and thus we obtain
\begin{align}\label{NaLim}
Na^D=\mathrm{const},\quad N\to\infty.
\end{align}

Note that the probability density $P(r)$ cannot be normalized to one \emph{after} this limit, because the integral $\int_{0}^{\infty}dr\, P(r)$ diverges at the lower limit of integration. This is a generic feature of dense packing of a finite volume in arbitrary dimension.

The structure factor always has the spike in the vicinity of zero momentum because of $S(\textbf{q}=0)=N$, but its behaviour is different in the different limits. In the limit of dense packing ($N \to\infty$ and the volume remains finite), the localization of the spike in the momentum space is equal to $(2\pi)^d/V$, and it does not shrink because $V= \mathrm{const}$. By contrast, in the usual thermodynamic limit this area shrinks and degenerates to a point when $N\to\infty$ and $N/V=\mathrm{const}$. This implies the $\delta$-function contribution to the structure factor at $\textbf{k}=0$. The $\delta$-function in Eq.~(\ref{relg_S_M}) leads to a constant shift $1$ of the pair distribution function: $g(\textbf{r})= 1 +\frac{1}{(2\pi)^2n} \int d^2k\, e^{i\textbf{k}\cdot\textbf{r}} [S(\textbf{k})-1]$, and we arrive at the standard relation between the pair distribution function and the structure factor. So, after the thermodynamic limit, we actually subtract the $\delta$-function and look at the behaviour of the residual part of the structure factor.

In principle, the structure factor can be extracted from the small-angle scattering data. We consider the scattering from a dilute system of many randomly oriented and located clusters, which are embedded into a homogeneous solid matrix. The area of each cluster is the square of size $s$ with the $N$ compactly packed disks inside. Then, in the case of the point-like densities located in the disk centers, the total intensity should indeed be $S(q=0)=N$ at $q=0$.

\section{The structure factor, mass-radius relation and pair distribution function for a set of dense segments}
\label{sec:entireDim}

We find the structure factor of non-overlapping segments of different lengths on a line, which are randomly mixed and then packed into a completely dense system. In accordance with the definition (\ref{Sqdef}), the structure factor in one dimension reads
\begin{align}
S(q) =\mathrm{Re}\big[ \omega(q)\big] +1,\label{SqGen}
\end{align}
where $\omega(q) =\frac{2}{N}\sum_{i<j}e^{i qr_{ij}}$. The distance between the centers of $i$th and $j$th segments is given by $r_{ij}=\frac{x_{i}}{2} +x_{i+1} +\cdots+x_{j-1} +\frac{x_{j}}{2}$. Here the lengths of the sequence of segments are denoted as  $x_{1},\cdots,x_{N}$.

Due to the randomness, the distribution of segment lengths are independent and described by the probability $P(x_{1}) \cdots P(x_{N})$ with $P(x)$ being the distribution function of a segment length, normalized to one: $\int_{0}^{\infty}dx\,P(x)=1$. For a finite numbers of segments, the probability is cut from below at $x=2a$, as discussed above in Sec.~\ref{sec:Distr_pack}. Note that radius in one dimension is equal to one-half of segment length.

We calculate $\omega(q)$ by integrating over the total distribution $P(x_{1}) \cdots P(x_{N})$. The mean value of a single term in the sum is given by $P^{2}(q/2)P^{j-i-1}(q)$, where $P(q)=\int_{0}^{\infty}dx\,P(x)e^{iqx}$ is the Fourier transform of $P(x)$, that is, its characteristic function. By adding up all the terms, we arrive at
\begin{align}
\omega(q) = 2P^2\left(\frac{q}{2}\right)\frac{P^{N}(q)-1-N\big[P(q)-1\big]}{N\big[P(q)-1\big]^2}. \label{omq}
\end{align}
This equation is general and takes into account the finite-size effects. Note that $P(q)\to 1$ when $q\to0$ due to the normalization, and we obtain from Eq.~(\ref{omq}) that $\omega(0)=N-1$, and hence $S(0)=N$, as it should be.

In the particular case of the power law, the distribution function of segment lengths is given by
\begin{align}
P(x) =
\frac{(2R)^D}{\Gamma(-D,a/R)x^{1+D}}\times\begin{cases}
e^{-x/(2R)}, &\text{for}\ 2a\leqslant x,\\
0, &\text{for} \ x < 2a
\end{cases}
\label{eq:PxSch}
\end{align}
with $\Gamma(s,x) =\int_{x}^{\infty}dt\,e^{-t} t^{s-1}$ being the incomplete gamma function. Here we replace the upper cut at $x=2R$ by the exponential, which is  more convenient technically.

The total number of segments whose radii exceed $r$ is determined by the cumulative distribution $N(r) = \frac{\Gamma\left(-D,r/R\right)}{\Gamma(-D,1)}$, normalized by the condition $N(R)=1$. Then the smallest radius $a$ is related to the total number of segments
\begin{align}\label{Na}
 N=N(a)=\frac{\Gamma(-D,a/R)}{\Gamma(-D,1)}\simeq \frac{R^D}{D \Gamma(-D,1)a^D},
\end{align}
where the last equality is the main asymptotics when $a\to0$. The total length of $N$ segments can be calculated as the average
$L_{N}=\langle x_1+\cdots+x_N\rangle =N\langle x\rangle$, which is given by
\begin{align}\label{LN}
L_{N}=N\int_{0}^{\infty}dx\,x P(x)=2R\frac{\Gamma(1-D,a/R)}{\Gamma(-D,1)}.
\end{align}
In the limit of infinite number of segments, the total length remains finite: $L =2R\,{\Gamma(1-D)}/{\Gamma(-D,1)}$.
Here $\Gamma(s)=\Gamma(s,0)$ is the ordinary gamma function.

The characteristic function of the distribution (\ref{eq:PxSch}) is easily obtained
\begin{align}\label{PqSch}
P(q)=\frac{(1+2qRi)^D\Gamma(-D,a/R+2qai)}{\Gamma(-D,a/R)}.
\end{align}
In the limit (\ref{NaLim}) of infinite number of segments, we find
\begin{align}\label{eq:w}
N[P(q)-1]\to w=\frac{\Gamma(1-D)[1-(1+2qRi)^D]}{D\Gamma(-D,1)}.
\end{align}
We note that $\Gamma(1-D)>0$ and $\Gamma(-D,1)>0$ when $0<D<1$. In the limit, Eq.~ (\ref{omq}) takes the form
\begin{align}\label{omN}
{\omega(q)}\simeq{N}\frac{2}{w^2}\big(e^{w}-1-w\big).
\end{align}
This result is also general as well as Eq.~(\ref{omq}).

At $q=0$, Eqs.~(\ref{SqGen}) and (\ref{omN}) yield ${S(q)}/{N}\simeq 1$, while at large values of wave vectors $q\gg q_\mathrm{c}$, we find
\begin{align}\label{Sqas}
 S(q)\simeq 1+\frac{2D\,\Gamma(-D,1)}{\Gamma(1-D)}\frac{\cos(\pi D/2)}{(2qR)^D}.
\end{align}
Here the crossover point $q_\mathrm{c}$ is introduced as
\begin{align}\label{qas}
q_\mathrm{c}=\frac{1}{2R}\left[\frac{|\cos(\pi D)|}{\cos(\pi D/2)}\right]^{1/D}.
\end{align}

\begin{figure}[!tb]
\centerline{\includegraphics[width=\columnwidth]{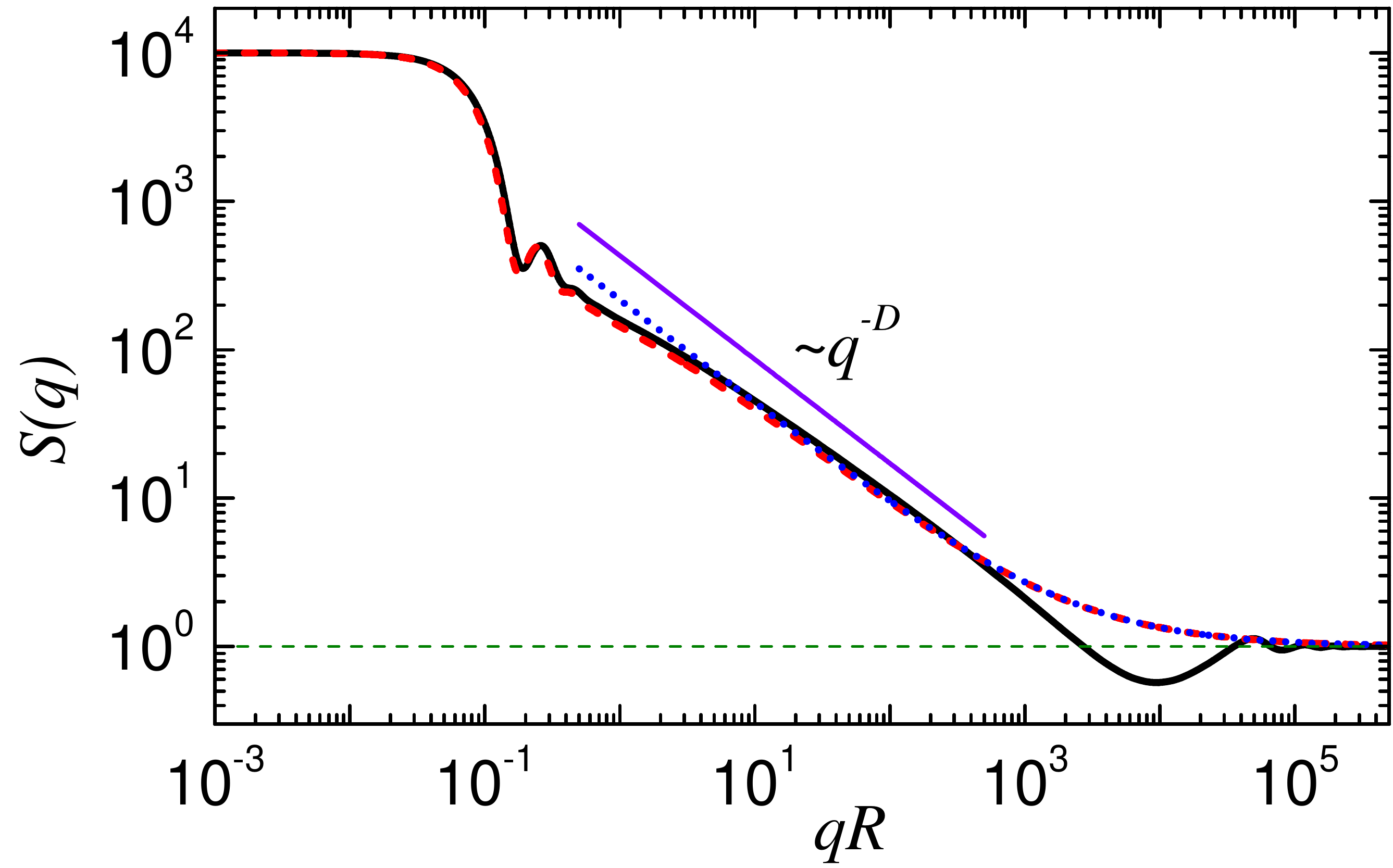}}
\caption{\label{fig:Sq1D} Theoretical curves for the structure factor vs. wave vector (in units of the inverse largest radius $1/R$).  Equation (\ref{SqGen}), together with Eqs.~(\ref{omq}), (\ref{Na}), and (\ref{PqSch}), takes into consideration the finite-size effects; it is shown in black solid line. The total number of segments $N=10000$ and the exponent of the power-law distribution $D=0.7$. The main asymptotics in $N$, given by Eqs.~(\ref{SqGen}), (\ref{eq:w}) and (\ref{omN}), is represented in red dashed line, and its long-range asymptotics (\ref{Sqas}) is indicated by blue dotted line. The crossover point (\ref{qas}) between the intermediate and asymptotic regimes is of order of $q_\mathrm{c}R \simeq0.7$ for the chosen control parameters, see Eq.~(\ref{qas}).
 }
 \end{figure}

Figure~\ref{fig:Sq1D} shows the structure factor $S(q)$ for a given set of control parameters. We observe that $S(q) \simeq 1$ when $q \rightarrow \infty$ and $S(q) = N$ at $q = 0$, as expected (see Sec.~\ref{sec:MrSqGen}). Note that the pair distribution function is equal to zero at $r=0$ due to the finite-size effects, as discussed in Sec.~\ref{sec:MrSqGen}. It follows form Eq.~(\ref{relg_S_M}) that $\int_{0}^{\infty}dq\,[S(q)-1]=0$. For this reason, there should always be a region where $S(q)<1$.

Once the structure factor is known, the mass-radius relation is obtained from Eq.~(\ref{relM_S}) for $d=1$. We compare the theoretical results for the mass-radius relation $M(r)$ with numerical simulations for $N=10000$ segments. The segments of sizes $x_{l}$ for $l=1,\cdots,N$ are generated with the cumulative distribution $\Gamma\left(-D,x_{l}/(2R)\right)/\Gamma(-D,1)=l$ and then randomly shuffled. The number of trials is equal to 20. For each trial, Eq.~(\ref{Mrdef}) is used for evaluating the mass-radius relation.

\begin{figure}[!tb]
\centerline{\includegraphics[width=\columnwidth]{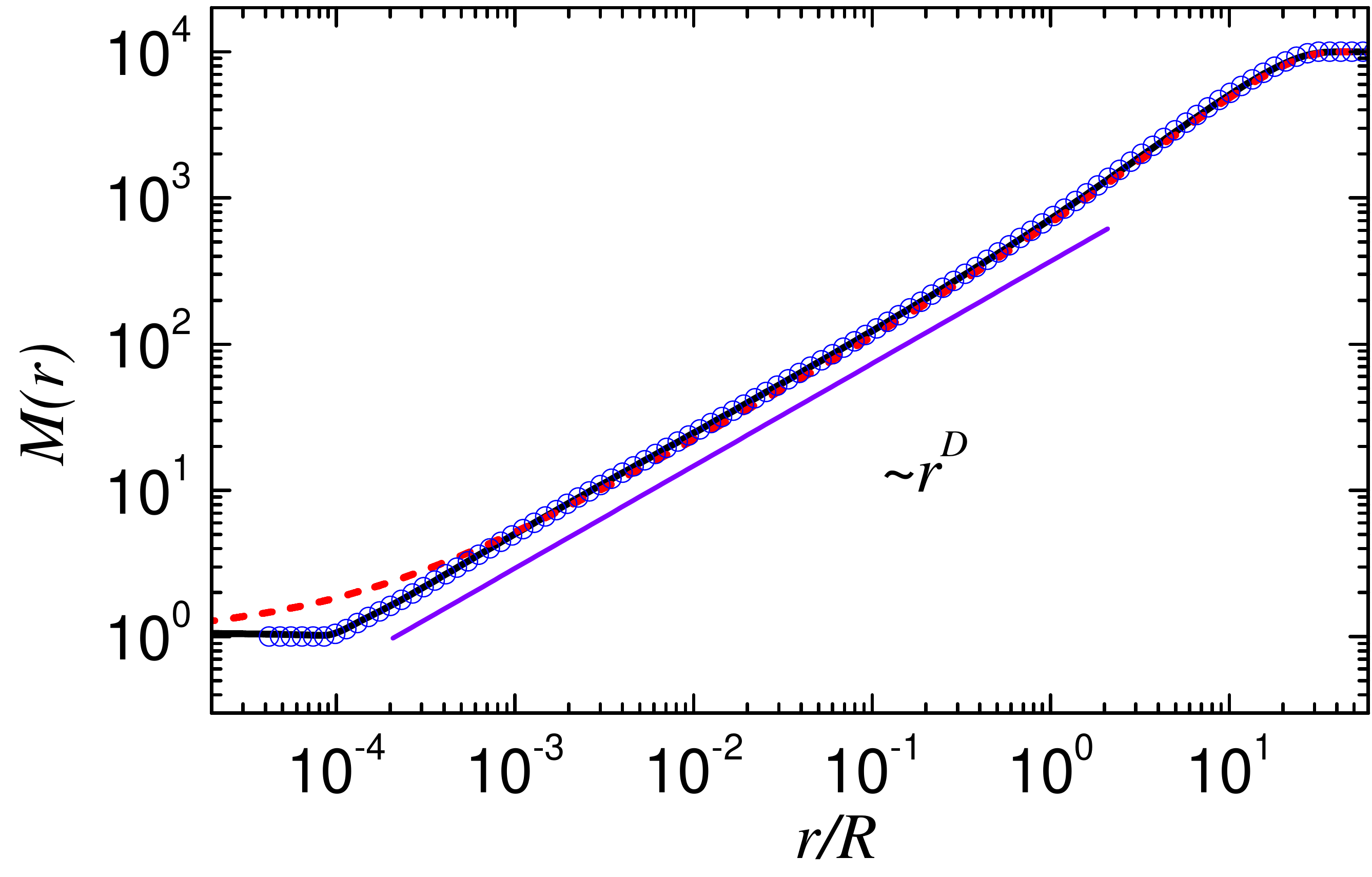}}
\caption{\label{fig:Mr1D}
The mass-radius relation for the centers of densely packed segments, whose lengths are randomly distributed according to the power-law with $D = 0.7$, Eq.~(\ref{eq:PxSch}). The distance is in units of the largest radius $R$. Open blue circles are the result of numerical simulations with N = 10000 segments. The theoretical curves are obtained by integration (\ref{relM_S}) of the structure factor for the finite number of segments  (black solid line) and the structure factor in the main asymptotics in $N$ (red dashed lines), see Fig.~\ref{fig:Sq1D}. Error bars, representing the standard deviations and corresponding to 20 trails of numerical calculations, are smaller than the diameters of the circles and thus not shown. The thin violet line is $r^D$ up to a factor. The borders of the fractal range (\ref{frr}) are of order of $r_\mathrm{min}/R=2a/R \simeq 8.4 \times 10^{-5}$ and $r_\mathrm{max}/R ={2\pi}/{(q_\mathrm{c}R)} \simeq 8.6$.
 }
\end{figure}

Figure~\ref{fig:Mr1D} shows very good agreement between the theoretical and numerical results. The upper plateau of $M(r)=N$ begins from the maximal distance between the segment centers, which is practically equal to the total length of the segments (\ref{LN}) $L_{N}/R\simeq 32.2$. The lower plateau $M(r) = 1$ extends up to the minimal distance between the segment centers, which is $2a$ with a good accuracy (see the discussion in Sec.~\ref{sec:MrSqGen} above). Between the plateaus, the mass-radius relation exhibits the power law $r^{D}$. The fractal range in real space is given by
\begin{align}\label{frr}
{2a}\lesssim {r} \ll \frac{2\pi}{q_\mathrm{c}},
\end{align}
where $q_\mathrm{c}$ is given by Eq.~(\ref{qas}). In the range ${2\pi}/{q_\mathrm{c}}\lesssim r \lesssim L_{N}$, the mass-radius relation deviates from $r^{D}$ and is nearly proportional to $r$. This behaviour looks like a reminiscence of the asymptotics of mass-radius relation for a large thermodynamic system (see Sec.~\ref{sec:MrSqGen} above).

\begin{figure}[!tb]
\centerline{\includegraphics[width=\columnwidth]{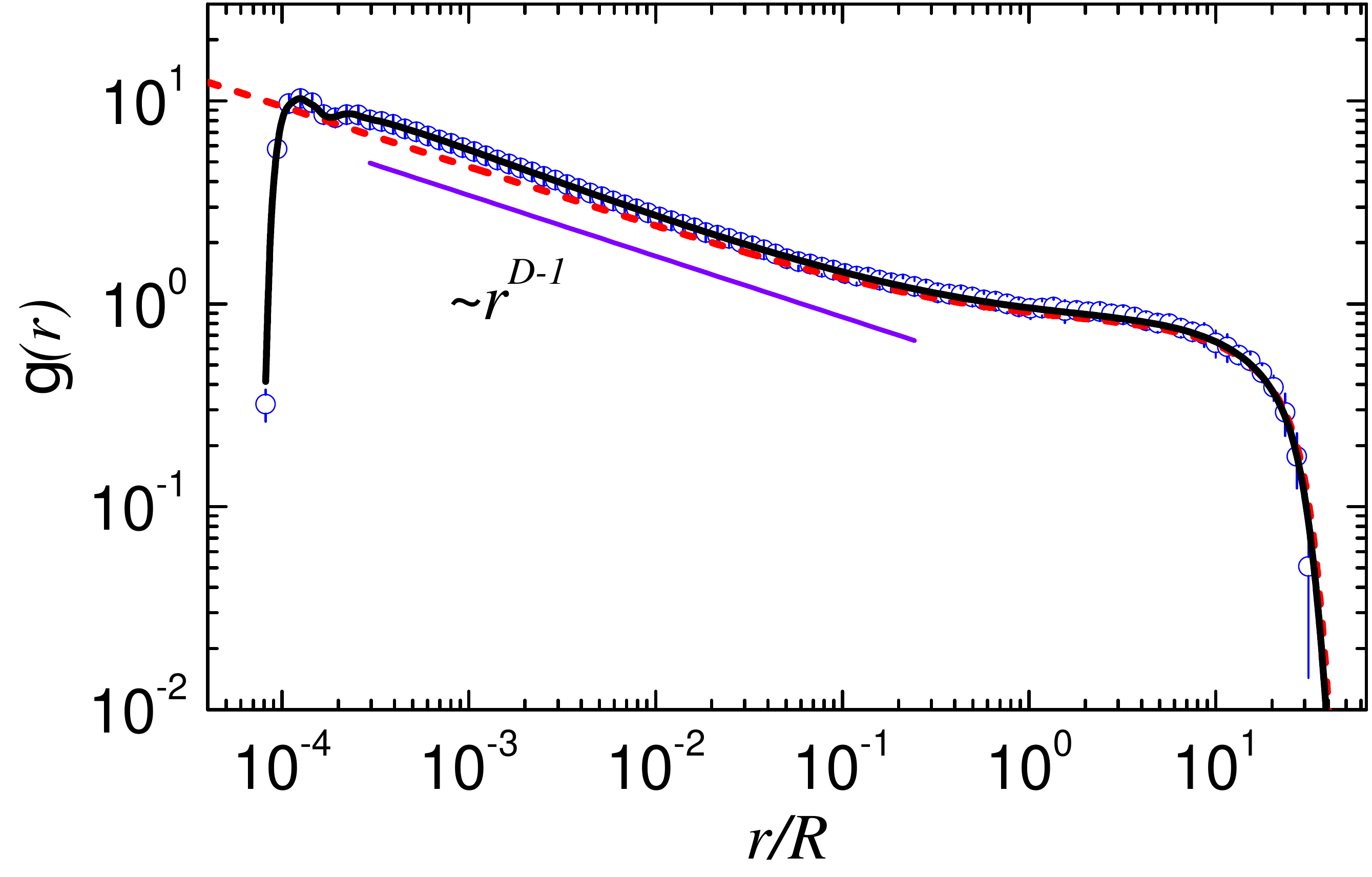}}
\caption{\label{fig:gr1D}
The pair distribution function for the centers of densely packed segments vs. distance. The units and control parameters are the same as in Fig.~\ref{fig:Mr1D}.  The numerical simulations (open blue circles) are obtained through the derivative $dM/dr$, Eq.~(\ref{relg_S_M}). The theoretical curves are calculated by integration of the structure factor, Eq.~(\ref{relg_S_M}), for the finite number of segments  (black solid line) and the infinite-number asymptotics (red dashed lines), see Fig.~\ref{fig:Sq1D}. Error bars show the standard deviations for 20 trials. The thin violet line is $r^{D-1}$ up to a factor.
 }
\end{figure}

The pair distribution function is shown in Fig.~\ref{fig:gr1D}. The theoretical and numerical values are obtained with the help of Eq.~\eqref{relg_S_M} by integrating $S(q)$ and taking the derivative of $M(r)$, respectively. The control parameters are the same as those used in Figs.~\ref{fig:Sq1D} and \ref{fig:Mr1D}. The agreement between the curves is very good, and their behaviour shows that $g(r) \propto r^{D-1}$ within the same fractal range as $M(r)$. In the range ${2\pi}/{q_\mathrm{c}}\lesssim r \lesssim L_{N}$, the pair distribution function is close to a constant, in accordance with the behaviour of $M(r)$ within the same range (see Fig.~\ref{fig:Mr1D}).

\section{The 2D densely packed disks}
\label{sec:2d}

\subsection{Model}
\label{sec:2Dcircles}
Similar to the one-dimensional case, we find the correlation properties of a system consisting of non-overlapping and randomly placed disks with radii following a power-law distribution with the exponent $1 < D < 2$. Here it is more convenient to use not exponential but direct cut of the distribution from above. We consider the following cumulative distribution
\begin{align}\label{Nr2d}
N(r)=\begin{cases}
0, &r\geqslant R,\\
(R/r)^D,& a\leqslant r<R,\\
N,&r\leqslant a,
\end{cases}
\end{align}
where $N(r)$ is the number of disks whose radii exceed $r$. Then we build a set of disk of radii $r_{i}$ that obey the inequalities $s/2 \geqslant R=r_{1} > \cdots > r_{N}=a$ and follow the distribution (\ref{Nr2d}):
\begin{equation}
r_{i} = R\ i^{-1/D},\quad\mathrm{for}\ \ i=1,\cdots, N.
    \label{eq:2dpl}
\end{equation}
Here $s$ is the edge length of a square, $a$ and $R$ are the smallest and largest radii, respectively. The total number of disks is given by $N=(R/a)^D$.

Then the disks should be arranged into the square without overlapping in accordance with the following algorithm. The center of the largest disk is randomly put inside the square until the whole disk is completely embedded in it. Then the same operation is repeated for the largest remaining disk, which is embedded into the residual free space in the square. The operation repeats until all the disks are inserted inside the square. Thus, this process corresponds to a random-sequential-addition packing~\cite{widom66}  in which the radii of disks follow a power-law distribution. For a finite $N$, the algorithm exhausts all possible positions of the $N$ non-overlapping disks inside the square.

The chosen sequence from largest to smallest radii influences the success of a given trial only. If we change the sequence (e.g. from smallest to largest) then too many trials would be unsuccessful (which means that we cannot put all the disks inside the square for a given trial), and the algorithm becomes too time-consuming.

The full compact packing is achieved in the limit $N \rightarrow \infty$, when the total area of the randomly placed non-overlapping disks is equal to the area of the square $s^2$. The initial radius $R$ determines the packing fraction, which is defined as the ratio $A_\mathrm{tot}/s^2$. The total area occupied by the disks is given by $A_\mathrm{tot} = \pi \sum_{i=1}^{N}r_{i}^{2} =\pi R^2\sum_{i=1}^{N}i^{-2/D} \simeq \pi R^2\zeta(2/D)$ with $\zeta(x)$ being the Riemann zeta function. Here the last equality is valid in the limit of the infinite number of disks. The area of the square $s^2$ should be greater than or equal to $A_\mathrm{tot}$, which gives us the upper limit of $R$:
\begin{equation}
R_\mathrm{max} = \frac{s}{\sqrt{\pi \zeta(2/D)}}.
    \label{eq:RsD}
\end{equation}
It follows from this equation that $R_\mathrm{max}<s/2$ for $1<D<2$, as it should be. The higher the exponent $D$, the lower $R_\mathrm{max}$. When $R$ exceeds $R_\mathrm{max}$, the algorithm cannot be completed for all $N$ disks provided $N$ is sufficiently large.

\begin{figure}[t]
\centerline{\includegraphics[width=0.65\columnwidth]{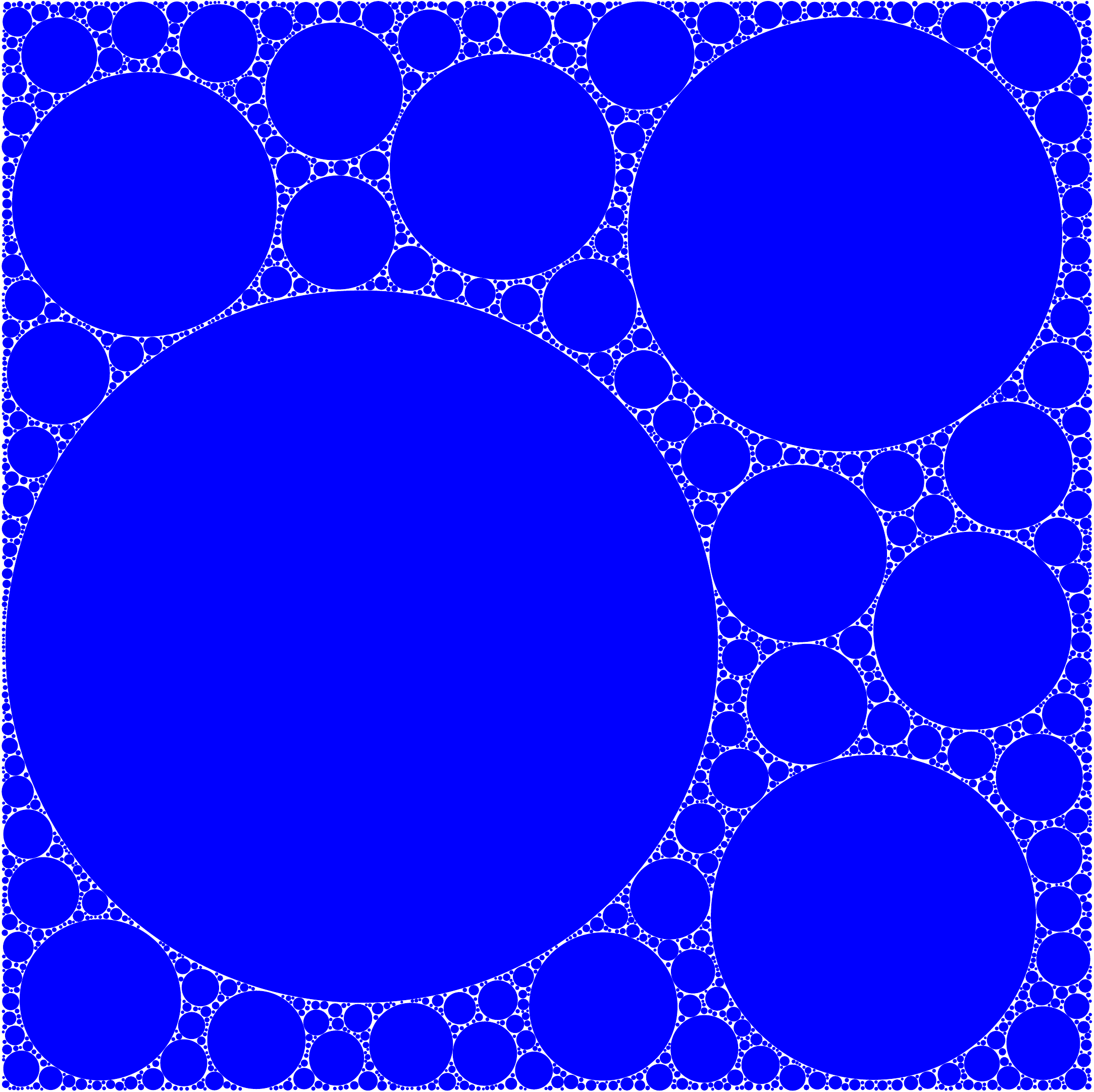}}
\caption{\label{fig:PL-model} Compact packing of a set of $N = 2407$ randomly distributed non-overlapping disks whose radii follow a power-law distribution with the exponent $D = 1.4$. The packing fraction (that is, percentage of the filled space) is about 0.96.
}
\end{figure}

Figure~\ref{fig:PL-model} shows a configuration of disks for the given control parameters. The ratio of $R_\mathrm{max} $ to $s$ is nearly equal to $0.329$ by Eq.~(\ref{eq:RsD}). We use $R/s = 0.327$, for which configurations with a high packing density are easily achieved.

 \subsection{Structure factor, mass-radius relation and pair distribution function}

\begin{figure}[!tb]
\centerline{\includegraphics[width=\columnwidth]{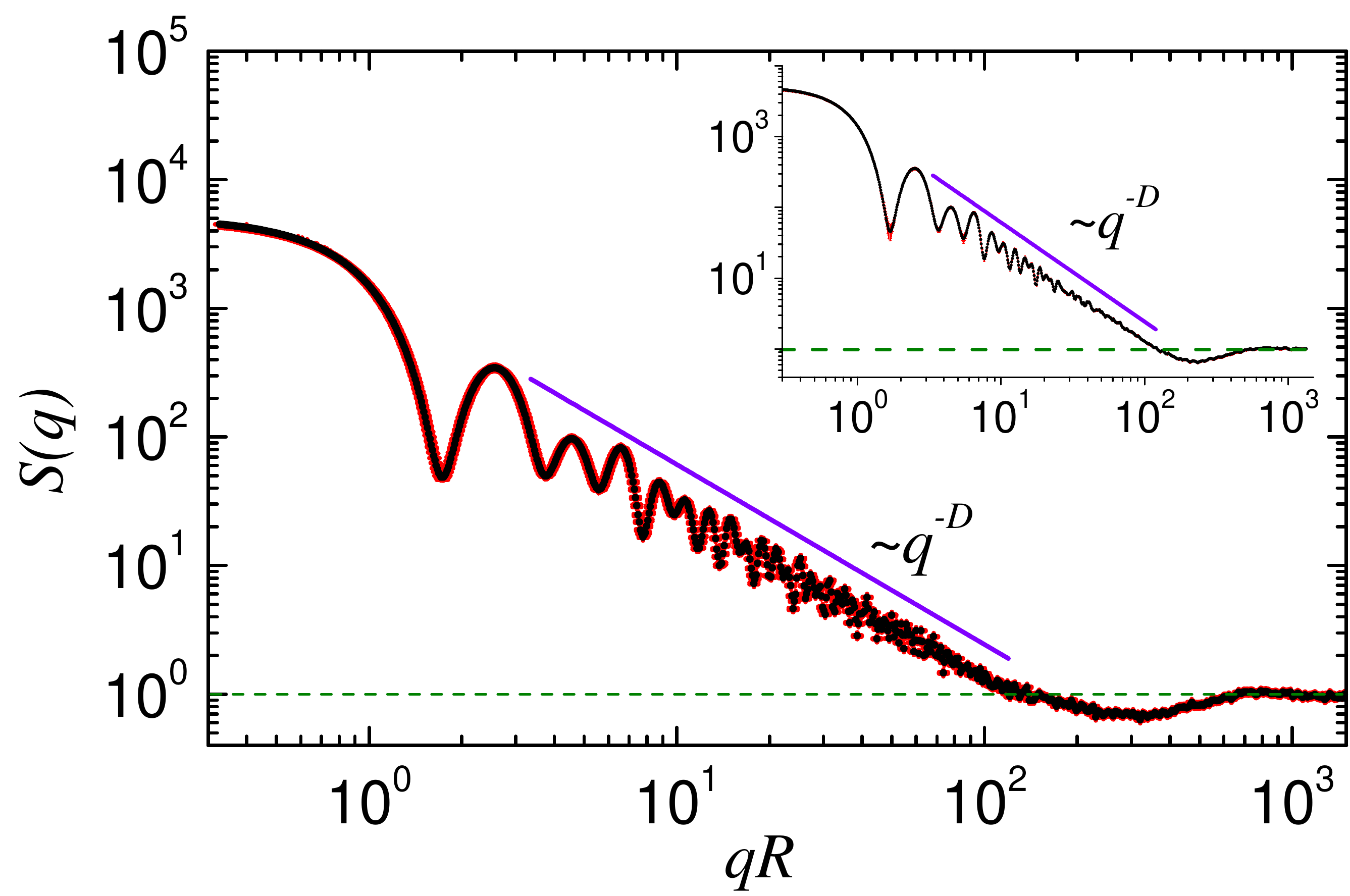}}
\caption{\label{fig:PL-Sq}
The structure factor \eqref{Sqdef} vs. wave vector (in units of the inverse radius $1/R$ of the largest disk). The control parameters are described in the main text, in particular, $s/R\simeq 3.1$. Black: average values over 20 trials. Red: the corresponding error bars representing the standard deviations. The violet line is $q^{-D}$ up to a factor. Inset: the structure factor \eqref{relS_M} calculated from the mass-radius relation shown in Fig.~(\ref{fig:Mr2D}) .
}
\end{figure}

\begin{figure}[!tb]
\centerline{\includegraphics[width=\columnwidth]{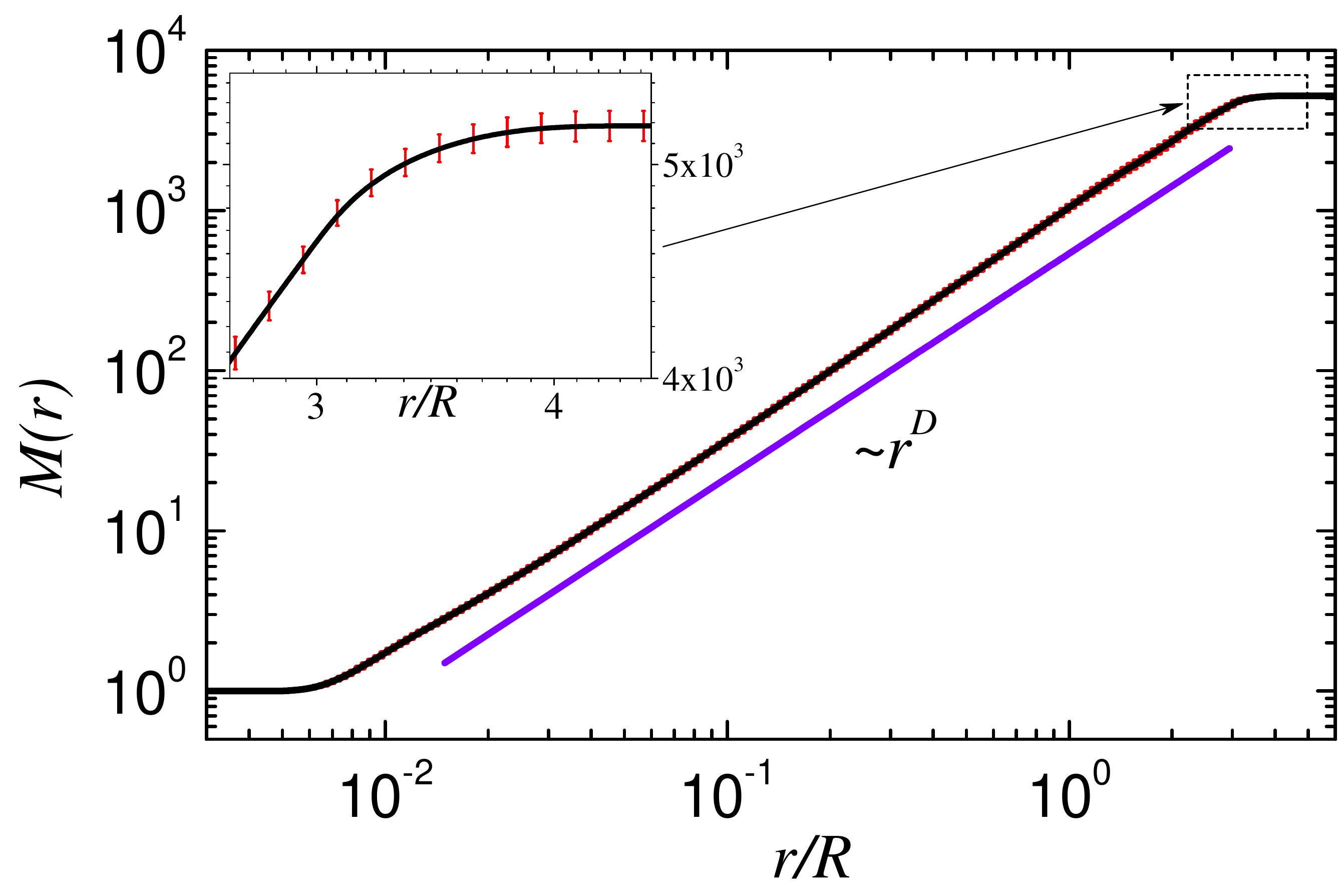}}
\caption{\label{fig:Mr2D}
The mass-radius relation for the centers of densely packed disks, whose radii are randomly distributed according to the power-law with the exponent $D=1.4$. The distance is in units of $R$. Black solid line: average of numerical simulations for 20 trials. Red bars: errors representing the standard deviations. The thin violet line is $r^D$ up to a factor. The borders of the fractal range are of order $r_\mathrm{min}/R = 2a/R \simeq 4.6\times 10^{-3}$ and $s/R \simeq 3.1$ (see the main text for details). The inset: $M(r)$ vs. $r/R$ near the upper knee of the curve. It shows that the mass-radius relation increases quite slowly in the range $s \lesssim r \lesssim \sqrt{2}s$.}
\end{figure}

When calculating the structure factor, mass-radius relation and pair distribution functions, we choose $N=5069$ and $D = 1.4$.  For these control parameters, the packing fraction is about 0.97, and $a/R \simeq 2.3\times10^{-3}$. By means of the above algorithm, 20 different configurations of disk positions are generated. For each trial, the mass-radius relation and structure factor are directly obtained with Eqs.~(\ref{Mrdef}) and (\ref{Sqdef}), respectively. Their average values and standard deviations are shown in Figs.~\ref{fig:PL-Sq} and \ref{fig:Mr2D}.

Let us discuss the behaviour of the structure factor in Fig.~\ref{fig:PL-Sq}. Except for the region near the first minimum (at $q\textcolor{blue}{s} \simeq 2\pi$), the errors sizes are negligible. This implies that at given control parameters $N$, $D$, and $R$, the structure factor is practically independent of a specific packing configuration provided the packing is sufficiently high. Similarly to the 1d case, within the fractal region $2\pi/R \lesssim q \ll 2\pi/a$, the structure factor decays proportional to $q^{-D}$. The lower and upper borders of the fractal region are related to the largest, and respectively smallest radii in the configuration. The asymptotics of the structure factor are the same as in 1d case: $S(q) \simeq 1$ when $q \gtrsim 2\pi /a$ and $S(q) = N$ at $q = 0$, in agreement the definition \eqref{Sqdef}. The lengthy region where $S(q)<1$ appears due to the finite-size effects, as explained in Sec.~\ref{sec:entireDim} above.

We also check numerically the consistency between the structure factor and the mass-radius relation. The analytical formula relating them is obtained from Eq.~(\ref{relg_S_M}) by means of the inverse Fourier transformation. The result reads
\begin{align}\label{relS_M}
S(q)=\int_{0}^{\infty}dz\,M\left(\frac{z}{q}\right)J_{1}(z).
\end{align}
This equation enables us to find numerically $S(q)$ from the simulations of $M(r)$ (see below). The results are presented in the inset of Fig.~\ref{fig:PL-Sq}. The agreement between the main diagram and the insert is excellent.

Figure \ref{fig:Mr2D} shows the data of the numerical simulations of mass-radius relation for the centers of densely packed disks. Similarly to the 1d case of densely packed segments, we have $M(r)= 1$ when $r \leqslant 2a$, and the fractal region $M(r) \propto r^{D}$ when $2a \ll r \lesssim r_{\mathrm{max}}$. Here the upper fractal border $r_{\mathrm{max}}$ is nearly equal to square edge $s$. When the distance exceeds the size of diagonal of the square $\sqrt{2}s$, we get $M(r) = N$ as discussed in Sec.~\ref{sec:MrSqGen} above. This asymptotic is reached through a transition region $s \lesssim r \lesssim \sqrt{2}s$, in which $M(r)$ has yet a small increment (see the inset in Fig.~\ref{fig:Mr2D}). This is because the contribution of pair distances between $s$ and $\sqrt{2}s$ to the total number of the pairs are rather small.

\begin{figure}[!tb]
\centerline{\includegraphics[width=\columnwidth]{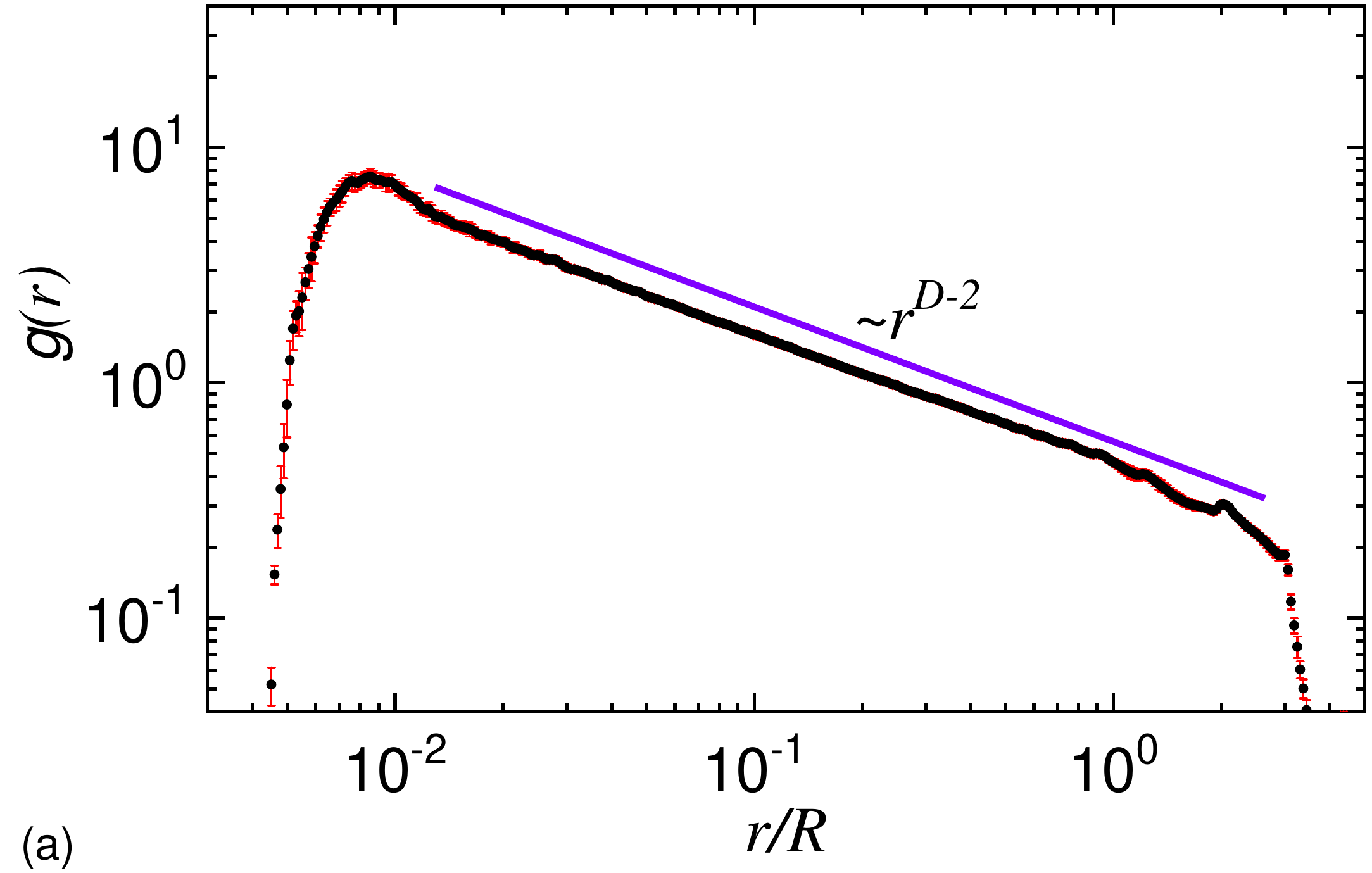}}
\centerline{\includegraphics[width=\columnwidth]{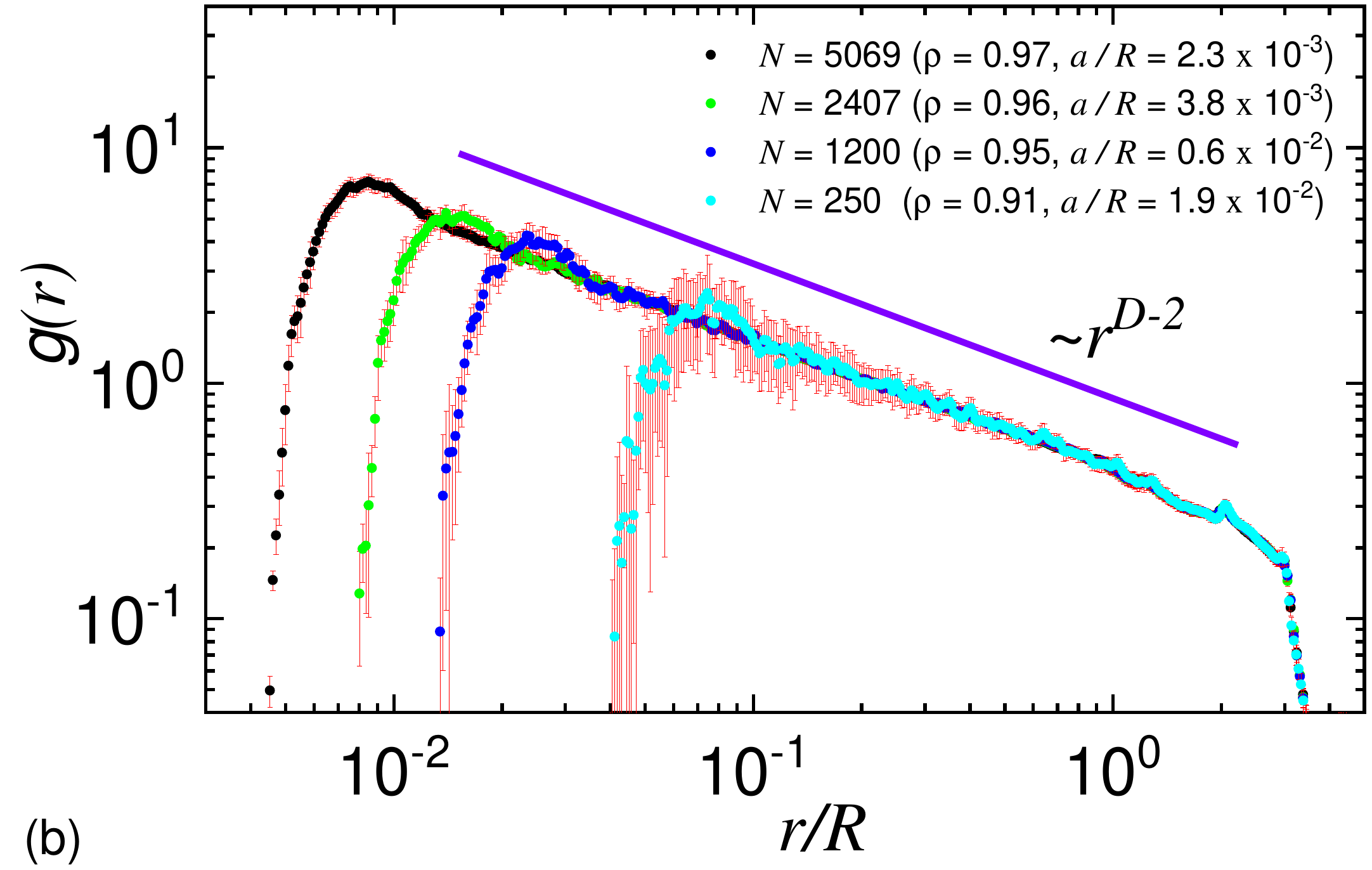}}
\caption{\label{fig:gr2D}
(a) The pair distribution function for the centers of densely packed disks vs. distance (in units of $R$). The numerical simulations (black circles) are obtained through the derivative of the mass-radius relation shown in Fig.~\ref{fig:Mr2D}, see Eq.~(\ref{relg_S_M}). The ratio $s/R$ is equal to $3.1$. (b) Variation of the pair distribution function with the number of particles $N$. Red bars: errors representing the standard deviations for 20 trials. The thin violet line is $r^{D-1}$ up to a factor.}
\end{figure}

The pair distribution function $g(r)$ is obtained through the derivative of $M(r)$, according to Eq.~\eqref{relg_S_M}. Its behaviour is represented in Fig.~\ref{fig:gr2D}(a), which shows that $g(r) \propto r^{D-2}$ in the fractal range. The errors within this range are also very small, similar to those in $M(r)$.

When the number of particles $N$ is varied, a similar behaviour of $g(r)$ is still observed [see Fig.~\ref{fig:gr2D}(b)]. The length of the fractal range is of order of the smallest radius $a$, which is determined by the total number of disks: $a=RN^{-1/D}$.
The smaller the $N$, the smaller the range and the higher the errors, and vice-versa. The increase of errors is due to decreasing of the packing fraction. As a result, many voids appear, and the positions of the disks with small radii becomes less correlated and, hence, the fluctuations increase.

\section{An approximate relation for the structure factor in arbitrary dimension}
\label{sec:empir}

In arbitrary dimension, one can derive a fitting analytical formula for the structure factor by means of a simple approximation for the mass-radius relation. The approximation assumes that $M(r)=[r/(2a\gamma)]^D$ within the range $2 a\gamma\leqslant r\leqslant 2R\gamma$ and it takes the constant values $1$ and $N$ for $r\leqslant 2a\gamma$ and $r\geqslant 2R\gamma$, respectively. Here $\gamma$ is the only dimensionless fitting parameter. The relation $(R/a)^D=N$ is supposed to be valid, as usual. It follows that $\frac{\partial M}{\partial r}=D r^{D-1}/(2a\gamma)^D$ within the range and zero elsewhere. Equation (\ref{relg_S_M}) tells us  that $\frac{1}{\Omega r^{d-1}}\frac{\partial M}{\partial r}$ is related to $S(q)-1$ through the Fourier transformation. Taking the inverse Fourier transformation yields
\begin{align}\label{Sqapp}
S(q)=1+N f_{D,d}(\gamma qaN^{1/D})-f_{D,d}\left(\gamma qa\right),
\end{align}
where we put by definition
\begin{align}
f_{D,d}(z)={}_{1}F_{2}\left(\frac{D}{2};\frac{d}{2},1+\frac{D}{2};-z^2\right)\nonumber
\end{align}
with  ${}_{1}F_{2}$ being the generalized hypergeometric function \cite{Slater66}.

The approximation formula (\ref{Sqapp}) is  a strongly oscillating function. A more realistic description requires smoothing to take into account an additional dispersion or experimental resolution. To this aim, without loos of generality, we consider a log-normal distribution of the overall length
\begin{equation}
D_{\mathrm{N}}(s) = \frac{1}{\sigma s (2\pi)^{1/2}}\exp\left( -\frac{[\log(s/\mu_{0})+\sigma^{2}/2]^{2}}{2\sigma^{2}} \right),
    \label{eq:DN}
\end{equation}
where $\sigma = [\log(1+\sigma_{\mathrm{r}}^{2})]^{1/2}$. The quantities $\mu_{0}$ and $\sigma_{\mathrm{r}}$ are the mean length and relative variance, i.e. $\mu_{0} \equiv \langle s \rangle_{D}$ and $\sigma_{\mathrm{r}} \equiv \left(\langle s^{2} \rangle_{D} - \mu_{0}^{2} \right)^{1/2}/\mu_{0}$ and $\langle \cdots \rangle \equiv \int_{0}^{\infty} \cdots D_{\mathrm{N}}(s)\mathrm{d}s.$

Then the smoothed structure factor is calculated as the average of Eq.~\eqref{Sqapp} over the distribution $D_\mathrm{N}$:
\begin{equation}
S_{\mathrm{sm}}(q) = \int_{0}^{\infty}S(q)  D_{\mathrm{N}}(s)\mathrm{d}s.
    \label{eq:polySF}
\end{equation}
Here we assume that all radii, including $a$ and $R$, are proportional to $s$.

\begin{figure}[!t]
\centerline{\includegraphics[width=.95\columnwidth]{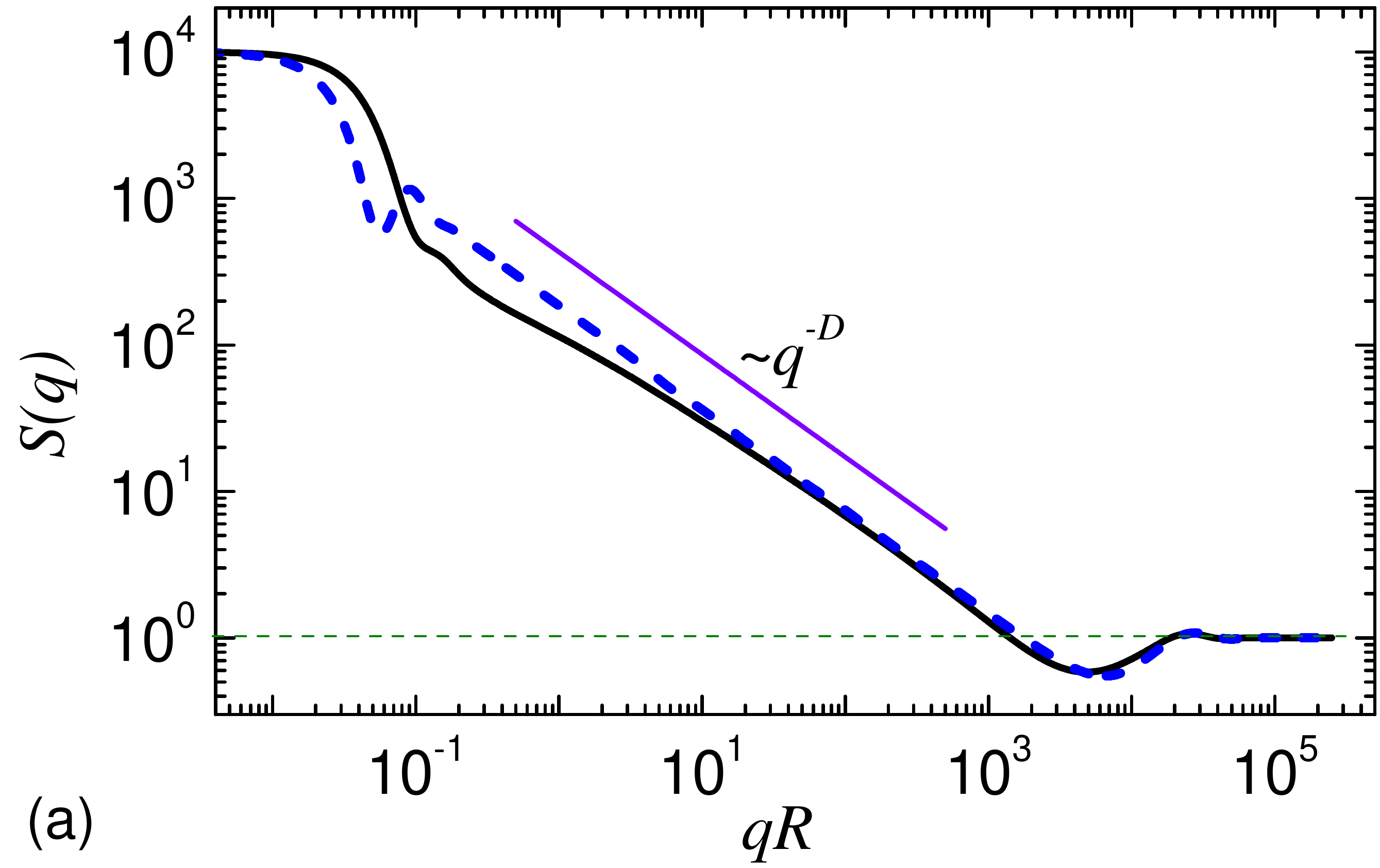}}
\centerline{\includegraphics[width=.95\columnwidth]{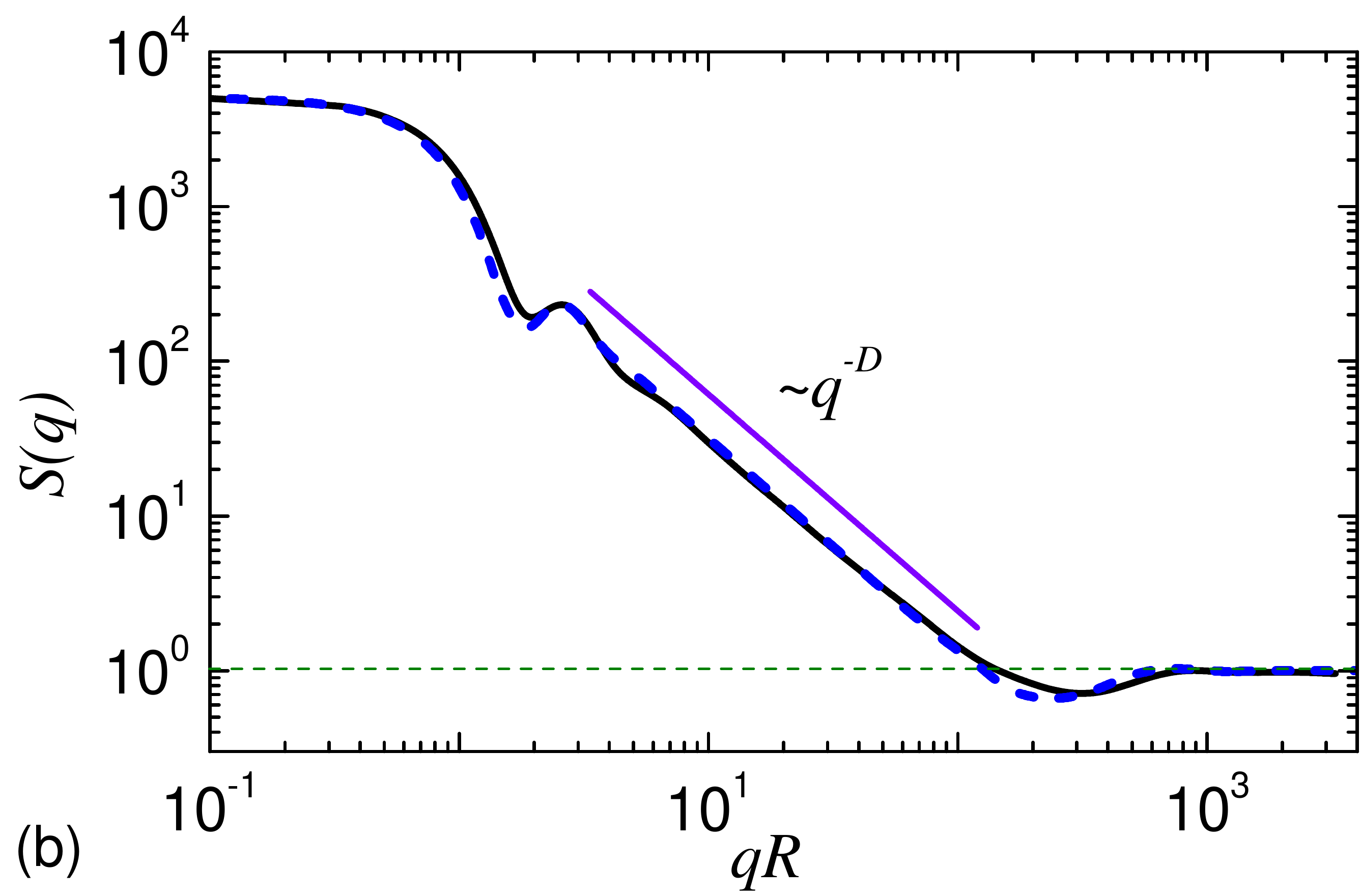}}
\caption{\label{fig:polySq1Dand2D}
The smoothed structure factor (\ref{eq:polySF}) vs. wave vector (in units of $1/R$) for a dense packing of segments (a) and disks (b) with the same control parameters as in Figs.~\ref{fig:Sq1D} and \ref{fig:PL-Sq}, respectively. Blue dashed line shows the smoothed approximate structure factor, given by Eqs.~(\ref{Sqapp})-\eqref{eq:polySF} with the relative variance $\sigma_{\mathrm{r}} = 0.2$. The fitting parameter $\gamma$ is chosen to be $1$ and $1.41$ in 1d and 2d, respectively. The thin violet line is $q^{-D}$ up to a factor. (a) Black solid line corresponds to the black line in Fig.~ \ref{fig:Sq1D} smoothed with equation (\ref{eq:polySF}). (b)  Black line: smoothed black line of Fig.~\ref{fig:PL-Sq}.
}
\end{figure}

The smoothed structure factor \eqref{eq:polySF} is represented in blue dashed line in Fig.~\ref{fig:polySq1Dand2D}. The power-law decay $S(q) \propto q^{-D}$ is recovered with the exponents $D$. Figure \ref{fig:polySq1Dand2D} also compares the theoretical and numerical polydisperse structure factors of 1d segments, smoothed by Eq.~\eqref{eq:polySF}. Note that the relation between $a$ and $R$ for the segments should be calculated with Eq.~(\ref{Na}), which involves the additional factor $D \Gamma(-D,1)$. As expected, the approximation \eqref{Sqapp} works better at large wave vectors, because the used approximation for the mass-radius relation is more precise at short distances. In both 1d and 2d cases, the agreement between the numerical and approximate structure factors is good in the fractal region.

Note that the suggested phenomenological relation (\ref{Sqapp}) can be used in more general cases than just a compact packing. When a mass-fractal aggregate is composed of the same particles, say, balls of equal radius, then the scattering intensity is given by the structure factor times the form factor of the ball, see, e.g., the discussion in Ref.~\cite{cherny11}. Thus, once experimental SAS data for the scattering intensity of the mass fractal and for the form factor of the basic units are available, then the structure factor of the mass fractal can be retrieved, and, one can treat Eq.~(\ref{Sqapp}) as a fitting formula for the small-angle scattering from mass-fractals formed by basic units of the same size.

The phenomenological relation like that was suggested in the early review by Teixeira \cite{Teixeira1988Small-angleSystems}, who used an exponential cut-off of the pair distribution function. By contrast, the equation (\ref{Sqapp}) is derived directly from the mass-radius relation with the cut-off at the edges of the fractal region. The sharp cut-off leads to oscillations and the smearing (\ref{eq:polySF}) is needed. The equation (\ref{Sqapp}) is valid in arbitrary dimensions and takes into consideration the finite-size effects. If the experimental SAS data include all the main scattering regions, i.e. the Guinier, fractal and Porod/asymptotic ones, then one can consider the fractal dimension, fractal edges and the number of basic units as possible fitting parameters.

The auxiliary parameter $\gamma$ is of order of one and it is difficult to interpret. In 1d and 2d, it is chosen to be $1$ and $1.41$, respectively (see Fig.~\ref{fig:polySq1Dand2D}), which suggests that $\gamma=\sqrt{d}$ in general. This hypothesis will be verified in subsequent publications.

\section{Conclusions}
\label{sec:concl}

In this paper, the structure factor $S(q)$, the mass-radius relation $M(r)$ and the pair distribution function $g(r)$ are used to explore the correlation properties of dense systems of packed particles with a power-law size distribution with the exponent $D$. It is shown that at high packing fraction, the correlation properties of densely packed and fractal systems are alike, that is, they exhibit the power-law behaviour in the fractal region: $S(q) \propto q^{-D}$, $M(r) \propto r^{D}$ and $g(r) \propto r^{D-d}$ (here $d$ is the dimension of space). The results are confirmed theoretically and numerically for 1d systems consisting of segments, and numerically for 2d systems consisting of disks. The finite-size effects are also studied and explained.

An important question about the limit $N\to\infty$ of fully dense packing is discussed in Sec.~\ref{sec:Distr_pack}. It is shown that for a power-law distribution it takes the form (\ref{NaLim}), which differs from the standard thermodynamic limit.

In 1d case, the analytical expression for the structure factor (\ref{SqGen}), in conjunction with Eq.~(\ref{omq}), is obtained not only for a power-law but for arbitrary distribution. It takes into consideration the finite-size effects.

An approximate formula \eqref{Sqapp} for the structure factor, valid in arbitrary dimension, is suggested. The parameters involved are the smallest radius $a$, the total number of $d$-dimensional spheres $N$, the power-law exponent $D$, and the fitting dimensionless parameter $\gamma$. The most relevant systems for which Eq.~\eqref{Sqapp} can be used involve objects compactly packed, and with centers having a much higher scattering-length density than that of the surrounding shells. The sizes of the objects should  follow a power-law distribution. A typical example would be biological macromolecules, for which the scattering-length density of the core is much higher as compared to the rest of the molecule.

The equations \eqref{Sqapp} and \eqref{eq:polySF} can also be used as a fitting formula for experimental small-angle scattering data from fractal aggregates that are composed of basic units of the same size (see the discussion at the end of Sec.~\ref{sec:empir}). This type of aggregates is common in aerosols and colloids and are formed via various processes, such as diffusion-limited cluster-cluster aggregation~\cite{sorensen10}.

As a prospect, the theory developed here for 1d systems can be generalized for two and three dimensions.

\section{Acknowledgements}
The authors acknowledge support from the JINR--IFIN-HH projects.

\bibliography{dp}

\end{document}